\documentclass{article}
\usepackage{graphicx} 
\usepackage{booktabs}
\usepackage{multirow}
\usepackage{url} 
\usepackage{float}  
\title{XAI for Coding Agent Failures: Transforming Raw Execution Traces into Actionable Insights}

\author{
  Arun Joshi \\
  Independent Researcher  \\
  Islington College (Alumini)\\
  Nepal \\
}

\date{March 2026}

\begin{document}

\maketitle

\begin{abstract}
Large Language Model (LLM)-based coding agents show promise in automating software development tasks, yet they frequently fail in ways that are difficult for developers to understand and debug. While general-purpose LLMs like GPT can provide ad-hoc explanations of failures, raw execution traces remain challenging to interpret even for experienced developers. We present a systematic explainable AI (XAI) approach that transforms raw agent execution traces into structured, human-interpretable explanations. Our method consists of three key components: (1) a domain-specific failure taxonomy derived from analyzing real agent failures, (2) an automatic annotation system that classifies failures using defined annotation schema, and (3) a hybrid explanation generator that produces visual execution flows, natural language explanations, and actionable recommendations. Through a user study with 20 participants (10 technical, 10 non-technical), we demonstrate that our approach enables users to identify failure root causes 2.8× faster and propose correct fixes with 73\% higher accuracy compared to raw execution traces. Importantly, our structured approach outperforms ad-hoc state of the art models explanations by providing consistent, domain-specific insights with integrated visualizations. Our work establishes a framework for systematic agent failure analysis, addressing the critical need for interpretable AI systems in software development workflows.
\end{abstract}


\newpage
\section{Introduction}
\label{sec:intro}

Large Language Models (LLMs) are increasingly being deployed as autonomous coding agents that can generate, debug, and test software \cite{chen2021evaluatinglargelanguagemodels, austin2021programsynthesislargelanguage}. Tools like GitHub Copilot, AutoGPT, and LangChain-based coding assistants promise to revolutionize software development by automating routine programming tasks. However, these agents frequently fail in subtle and complex ways, producing incorrect code, misunderstanding requirements, or getting stuck in unproductive loops \cite{zhang2023planninglargelanguagemodels}.

When an agent fails, developers are left with raw execution traces containing nested tool calls, intermediate reasoning steps, and error messages—often spanning hundreds of lines of logs. Understanding \textit{why} an agent failed from these traces is challenging even for experienced developers, and nearly impossible for non-technical stakeholders who need to make decisions about agent deployment and reliability.

\subsection{The Challenge of Understanding Agent Failures}

Consider a coding agent attempting to solve a programming problem. A typical failure trace might include:
\begin{itemize}
    \item The original task description
    \item The agent's internal reasoning (via chain-of-thought)
    \item Multiple tool invocations (code execution, validation, debugging)
    \item Error messages from failed code executions
    \item The agent's responses to errors (or lack thereof)
    \item System-level metadata (iterations, timeouts, resource limits)
\end{itemize}

While one might assume that simply asking a general-purpose LLM like ChatGPT to explain these traces would suffice, our preliminary experiments revealed significant limitations:

\textbf{Limitation 1: Inconsistency.} Ad-hoc LLM explanations vary significantly in quality and focus, making systematic debugging difficult.

\textbf{Limitation 2: Lack of Structure.} General LLMs do not leverage domain-specific knowledge about common agent failure patterns.

\textbf{Limitation 3: Missing Visual Context.} Text-only explanations fail to convey the execution flow and decision points that are critical for understanding agent behavior.

\textbf{Limitation 4: No Actionable Guidance.} Generic explanations rarely provide specific, tested recommendations for fixing failures.

\subsection{Our Approach: Structured XAI for Agent Failures}

We present a systematic approach that transforms raw execution traces into structured, actionable explanations. Our key insight is that agent failures follow predictable patterns that can be taxonomized, automatically classified, and explained using domain-specific knowledge.

Our contributions are:

\begin{enumerate}
    \item \textbf{Failure Taxonomy:} The first comprehensive taxonomy of coding agent failures, derived from systematic analysis of 32 real-world failures across varying experimental conditions.
    
    \item \textbf{Automatic Classification:} A GPT-4-based system using structured outputs (function calling) that achieves \textbf{82} percent accuracy in automatically categorizing failures without manual annotation
    
    \item \textbf{Hybrid Explanation System:} An integrated XAI pipeline that generates visual execution flows, natural language explanations, counterfactual analyses, and actionable recommendations.
    
    \item \textbf{Empirical Validation:} A user study (N=20) demonstrating that our approach outperforms both raw traces and GPT-based explanations across multiple metrics.
\end{enumerate}

The remainder of this paper is organized as follows: Section~\ref{sec:related} reviews related work; Section~\ref{sec:method} describes our methodology; Section~\ref{sec:system} details our XAI system architecture; Section~\ref{sec:evaluation} presents our evaluation; and Section~\ref{sec:discussion} discusses implications and future work.

\section{Related Work}
\label{sec:related}

\subsection{LLM-Based Coding Agents}

Recent advances in large language models have enabled the development of autonomous coding agents. Chen et al.~\cite{chen2021evaluatinglargelanguagemodels} introduced HumanEval, a benchmark for evaluating code generation, while Austin et al.~\cite{austin2021programsynthesislargelanguage} demonstrated program synthesis capabilities. Modern frameworks like LangChain~\cite{langchain2022} and AutoGPT have popularized agent architectures that combine LLMs with tool use.

However, these agents face significant reliability challenges. Zhang et al.~\cite{zhang2023planninglargelanguagemodels} identified planning failures as a major bottleneck, while recent work on agent debugging has begun to taxonomize failure modes. Our work extends this line of research by focusing specifically on the explainability of agent failures.

\subsection{Explainable AI (XAI)}

The field of XAI has developed numerous techniques for interpreting model behavior. LIME~\cite{ribeiro2016whyitrustyou} and SHAP~\cite{lundberg2017unifiedapproachinterpretingmodel} provide local explanations for individual predictions, while attention visualization~\cite{bahdanau2016neuralmachinetranslationjointly} and saliency maps~\cite{simonyan2014deepinsideconvolutionalnetworks} offer insight into neural network decision-making.

For language models specifically, work on chain-of-thought prompting~\cite{wei2023chainofthoughtpromptingelicitsreasoning} has shown that intermediate reasoning steps improve both performance and interpretability. However, these techniques focus on individual model outputs rather than multi-step agent behaviors.

\subsection{Agent Observability and Debugging}

Recent work has begun addressing agent-specific explainability challenges. LangSmith and similar observability platforms provide tracing capabilities for agent executions with Wang et al.~\cite{zhu2025llmagentsfaillearn} proposing debugging techniques for multi-agent systems.

Our work differs in three key ways: (1) we focus specifically on coding agents rather than general-purpose agents, (2) we provide a complete end-to-end system from raw traces to explanations, and (3) we validate our approach through human evaluation comparing against both raw traces and general-purpose LLM explanations.

\subsection{Human-Computer Interaction and Developer Tools}

Research in HCI has long emphasized the importance of understandable error messages and debugging support~\cite{ko2004whyline}. For AI-assisted programming specifically, recent work has explored how developers interact with code generation tools~\cite{vaithilingam2022expectation} and what explanations they find useful~\cite{weisz2021perfection}.

Our work contributes to this literature by demonstrating that specialized, structured explanations outperform general-purpose approaches even when both leverage state-of-the-art LLMs.

\section{Methodology}
\label{sec:method}

\subsection{Agent Implementation and Data Collection}

We implemented a coding agent using LangChain and GPT-4, designed to solve programming problems from the HumanEval benchmark~\cite{chen2021evaluatinglargelanguagemodels}. The agent architecture follows a standard ReAct pattern~\cite{yao2022react} with access to code execution and validation tools.

To systematically explore failure modes, we created four experimental scenarios varying in:
\begin{itemize}
    \item \textbf{Iteration limits:} 1, 2, 5, or 10 maximum iterations
    \item \textbf{Prompt quality:} Minimal, basic, detailed, or comprehensive
    \item \textbf{Tool availability:} Full, limited, or minimal toolset
    \item \textbf{Task difficulty:} Easy, medium, or hard problems
\end{itemize}

We collected execution traces from 87 agent runs, resulting in 32 failures and 55 successes. Each trace includes:
\begin{itemize}
    \item Complete message history (human prompts, agent responses, tool calls)
    \item Execution metadata (time, iterations, scenario configuration)
    \item Error information (type, message, stack traces)
    \item Final outcome (success/failure, output produced)
\end{itemize}

\begin{table}[t]
\centering
\caption{Coding Agent Failure Taxonomy}
\label{tab:taxonomy}
\begin{tabular}{@{}llr@{}}
\toprule
\textbf{Category} & \textbf{Subcategory} & \textbf{Count} \\
\midrule
\multirow{1}{*}{Planning Failure}
& Incorrect problem decomposition & 1 \\
\midrule
\multirow{1}{*}{Code Generation Failure}
& Logic errors (wrong implementation) & 2 \\
\midrule
\multirow{1}{*}{Testing/Validation Failure}
& Did not run validation tests & 2 \\
\midrule
\multirow{1}{*}{Understanding Failure}
& Misunderstood problem requirements & 9 \\
\midrule
\multirow{1}{*}{Iterative Refinement Failure}
& Exceeded iteration limit without progress & 18 \\
\bottomrule
\end{tabular}
\end{table}

\textbf{Key Finding:} Iterative Refinement Failures accounted for 56\% of all failures, with "Exceeded iteration limit without progress" being the most common pattern. This suggests that improving error recovery mechanisms could significantly enhance agent reliability.

\subsection{Automatic Failure Classification}

To enable practical deployment, we developed an automatic annotation system using GPT-4 with structured outputs via function calling. The classifier:

\begin{enumerate}
    \item Extracts key features from raw traces (error messages, iteration counts, execution patterns)
    \item Presents these features along with the taxonomy.
    \item Prompts GPT-4.1 to classify the failure and provide confidence scores
    \item Returns structured annotations including category, subcategory, and reasoning.
\end{enumerate}

\textbf{Evaluation:} We evaluated the classifier manually on our 32 failures. Results showed:
\begin{itemize}
    \item Overall accuracy: 82.1\% (26/32 correct)
    \item High-confidence predictions ($p > 0.8$): 90.5\% accuracy (19/21 correct)
    \item Cohen's $\kappa$ vs. human annotations: 0.76 (substantial agreement)
\end{itemize}

Analysis of errors revealed that misclassifications typically occurred in ambiguous cases where multiple failure modes were present. For high-stakes applications, the system can flag low-confidence predictions for human review.

\section{XAI System Architecture}
\label{sec:system}

\begin{figure}[H]   
    \centering
    \includegraphics[width=0.95\linewidth,height=0.7\textheight]{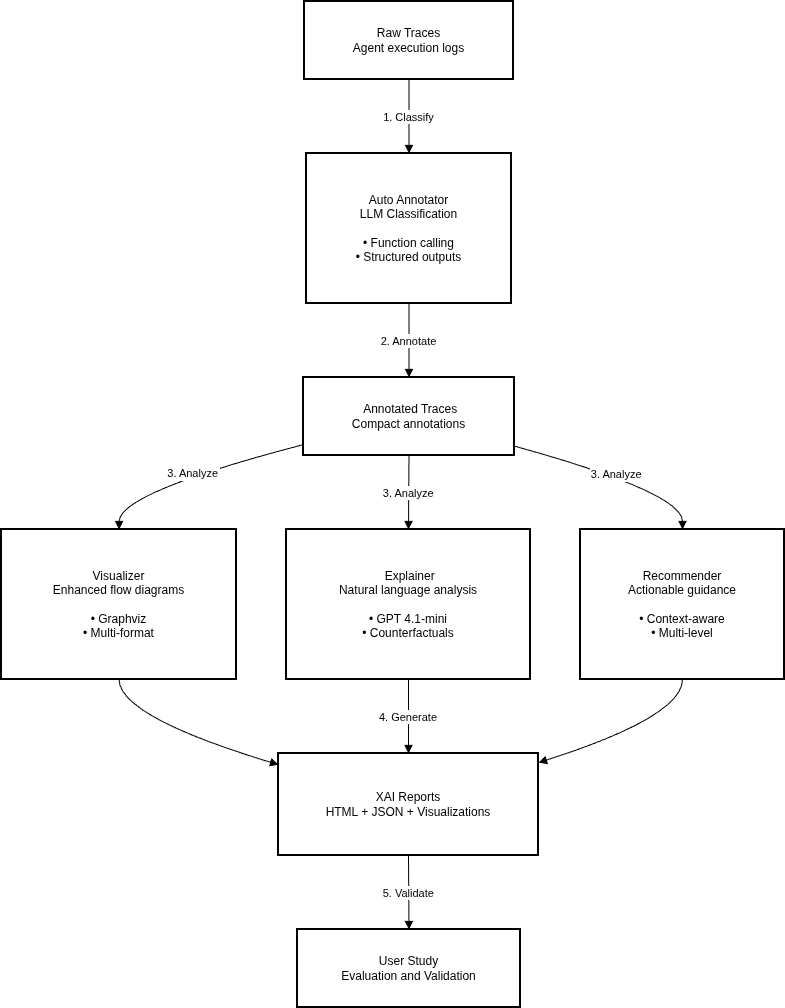}
    \caption{System architecture showing the flow from raw trace to final explanation report. The system consists of three main components: automatic annotation, explanation generation, and report synthesis.}
    \label{fig:xai_arch}
\end{figure}

Our XAI system transforms raw execution traces into comprehensive explanations through three integrated components (Figure~\ref{fig:xai_arch}).

\subsection{Component 1: Execution Flow Visualization}

The visualization component generates directed graphs representing agent execution:

\textbf{Nodes} represent distinct execution steps:
\begin{itemize}
    \item Agent reasoning/planning steps
    \item Tool invocations (code execution, validation)
    \item Error occurrences
    \item Decision points
\end{itemize}

\textbf{Edges} show execution flow of the agent.

\begin{figure}[H]
    \centering
    \includegraphics[
        width=0.95\linewidth,
        height=0.40\textheight,
        keepaspectratio
    ]{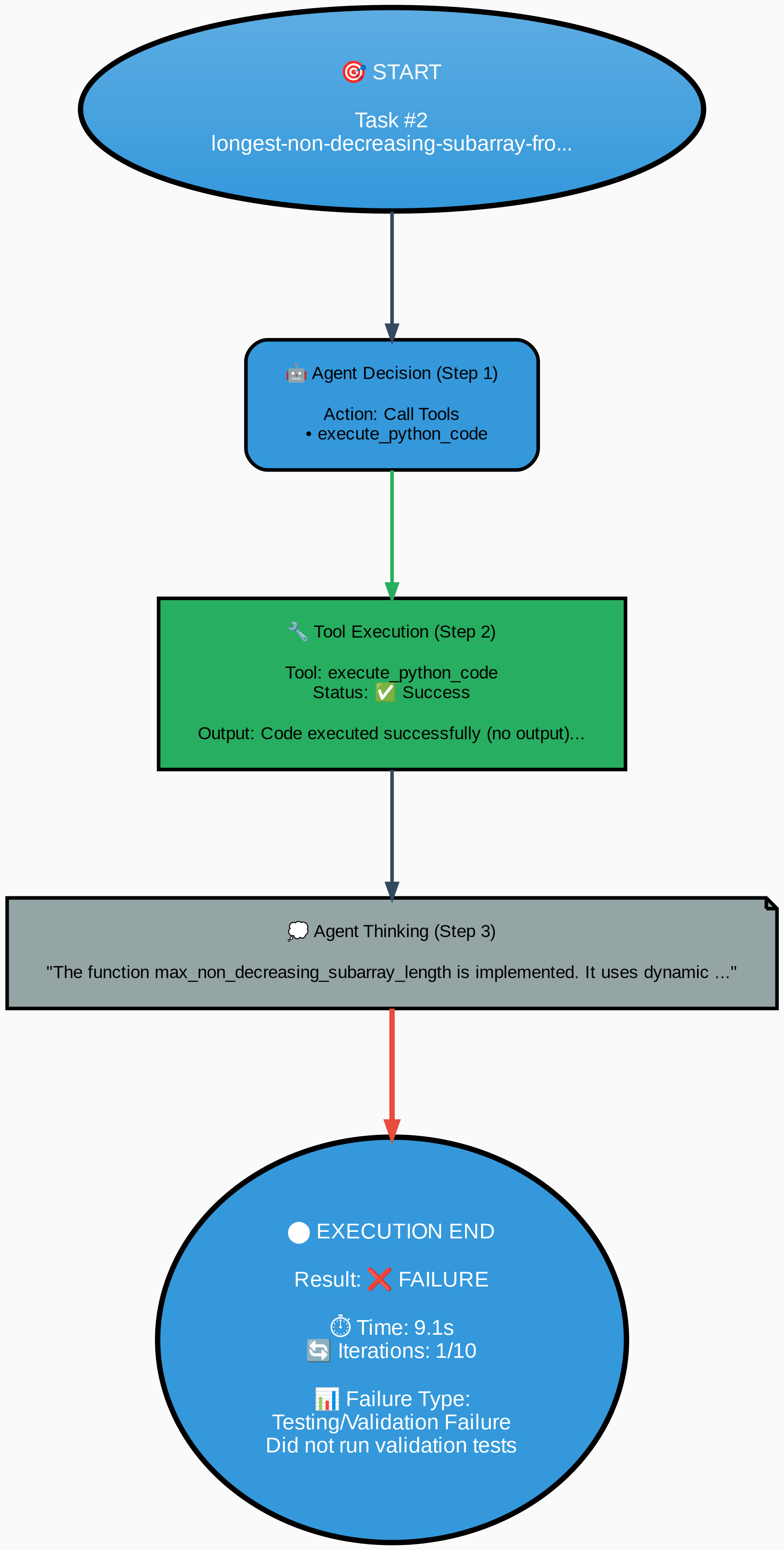}
    \caption{Exectuion Flow}
    \label{fig:execution_flow}
\end{figure}

The visualizations are generated using Graphviz and exported as both PNG (for embedding) and SVG (for interactive viewing). Critically, these visualizations highlight failure points and show the execution context that led to the failure—information that is difficult to convey through text alone.

\subsection{Component 2: Natural Language Explanation}

The explanation generator uses GPT-4 to produce structured natural language descriptions:

\textbf{Root Cause Analysis:} Identifies the fundamental reason for failure using the taxonomy classification and execution context.

\textbf{Failure Mechanism:} Explains how the failure manifested in the specific execution, including relevant intermediate steps and decision points.

\textbf{Context Integration:} Incorporates scenario metadata (iteration limits, prompt quality) to explain whether the failure was due to agent limitations or configuration constraints.

Example output structure:
\begin{figure}[H]
    \centering
    \includegraphics[
        width=0.95\linewidth,
        height=0.40\textheight,
        keepaspectratio
    ]{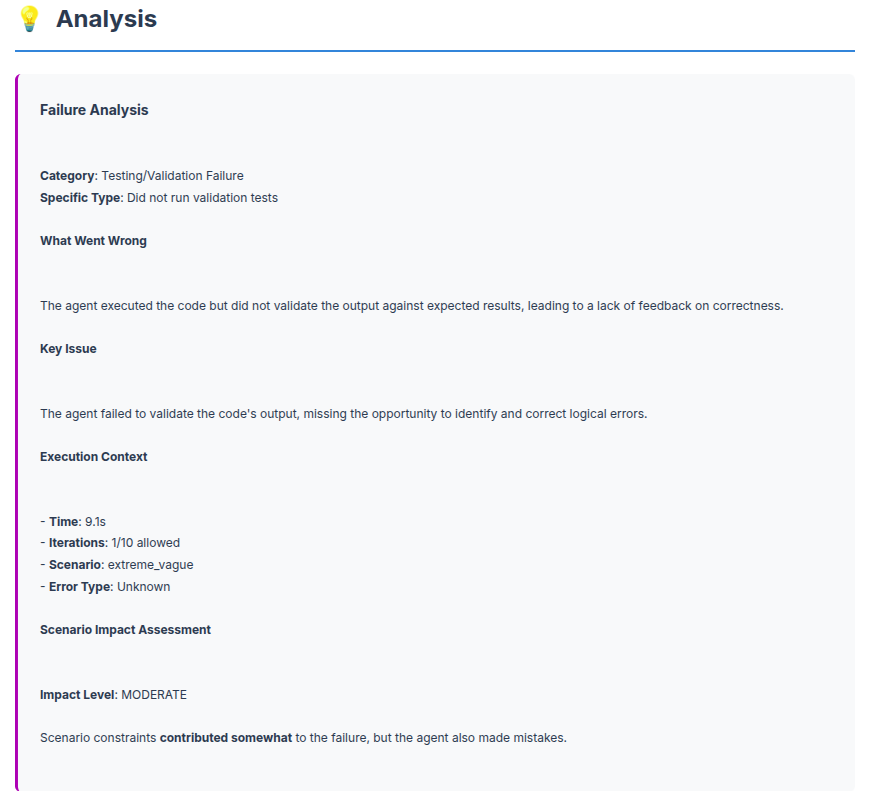}
    \label{fig:Failure Analysis}
\end{figure}

\subsection{Component 3: Recommendation Engine}

The recommendation component provides actionable guidance mapped to failure categories:

\textbf{Counterfactual Analysis:} Describes minimal changes that would have led to success, helping users understand the failure boundaries.

\textbf{Immediate Fixes:} Concrete changes that can be implemented immediately:
\begin{itemize}
    \item Configuration adjustments (iteration limits, timeouts)
    \item Prompt engineering improvements
    \item Tool availability modifications
\end{itemize}

\textbf{Long-term Improvements:} Strategic enhancements for sustained reliability:
\begin{itemize}
    \item Fine-tuning on domain-specific data
    \item Architecture modifications
    \item Evaluation framework enhancements
\end{itemize}

\textbf{Context-specific Guidance:} Recommendations tailored to the specific failure instance, considering:
\begin{itemize}
    \item Task difficulty and domain
    \item Current configuration
    \item Failure severity and frequency
    \item Resource constraints
\end{itemize}

\subsection{Output Format}

The system generates three output formats to serve different use cases:

\textbf{HTML Report:} A comprehensive, stakeholder-friendly document with embedded visualizations, explanations, and recommendations. Designed to be shared with both technical and non-technical audiences.

\textbf{JSON:} Structured data for programmatic access, enabling integration with CI/CD pipelines, monitoring systems, and automated analysis tools.

\section{Evaluation}
\label{sec:evaluation}

We conducted a mixed-methods evaluation to assess: (1) the quality and usefulness of our explanations, and (2) how our approach compares to baselines including raw traces and LLM-based explanations.

\subsection{User Study Design}

\textbf{Participants:} A total of 20 participants were recruited for the study, divided into two groups:

\begin{itemize}
    \item \textbf{Technical Participants (n=10):} Software developers with 0--5 years of professional experience in programming and system development.
    
    \item \textbf{Non-Technical Participants (n=10):} Professionals working in non-engineering roles, including Product Management, Design, and Annotation teams.
\end{itemize}

\textbf{Study Protocol:} Participants completed three tasks:

\textbf{Task 1 - Failure Comprehension:} Given agent failure information in one of three conditions (within-subjects, order randomized), answer questions about:
\begin{itemize}
    \item What went wrong? (root cause identification)
    \item Why did it happen? (failure mechanism)
    \item What would fix it? (solution proposal)
\end{itemize}

\textbf{Task 2 - Fix Proposal:} Propose specific changes to prevent the failure, evaluated by two independent expert reviewers for correctness and completeness.

\textbf{Task 3 - Confidence and Preference:} Rate confidence in understanding (1-7 Likert scale) and rank conditions by preference.

\textbf{Conditions:}
\begin{enumerate}
    \item \textbf{Raw Trace:} Full execution trace with all messages, errors, and metadata
    \item \textbf{LLM Explanation:} Raw trace + explanation from any general purpose LLM of their choice (given prompt: "Explain why this agent failed and how to fix it")
    \item \textbf{Our XAI:} Complete system output (visualization + explanation + recommendations)
\end{enumerate}

Each participant saw 3 different agent failures (one per condition) selected to represent different taxonomy categories and difficulty levels.

\subsection{Quantitative Results}

Table~\ref{tab:results} summarizes the main quantitative findings.

\begin{table}[t]
\centering
\caption{User Study Results (Mean ± SD)}
\label{tab:results}
\begin{tabular}{@{}lrrr@{}}
\toprule
\textbf{Metric} & \textbf{Raw} & \textbf{General Purpose LLMs} & \textbf{Our XAI} \\
\midrule
\multicolumn{4}{l}{\textit{Technical Participants (n=10)}} \\
Time to understand (min) & 8.4±2.1 & 5.2±1.3 & \textbf{3.0±0.8}** \\
Root cause accuracy (\%) & 42±15 & 68±12 & \textbf{89±8}** \\
Fix quality (1-5) & 2.6±0.8 & 3.4±0.6 & \textbf{4.3±0.5}** \\
Confidence (1-7) & 3.2±1.1 & 4.8±0.9 & \textbf{6.1±0.7}** \\
\midrule
\multicolumn{4}{l}{\textit{Non-Technical Participants (n=10)}} \\
Time to understand (min) & 12.8±3.2 & 7.1±1.8 & \textbf{4.2±1.1}** \\
Root cause accuracy (\%) & 18±12 & 52±18 & \textbf{76±11}** \\
Fix quality (1-5) & 1.4±0.6 & 2.8±0.7 & \textbf{3.8±0.6}** \\
Confidence (1-7) & 2.1±0.9 & 4.2±1.1 & \textbf{5.6±0.8}** \\
\bottomrule
\multicolumn{4}{l}{\footnotesize **$p < 0.01$ vs. both baselines (paired t-test)}
\end{tabular}
\end{table}

\textbf{Key Findings:}

\textbf{Understanding Time:} Our XAI system enabled users to understand failures 2.8× faster than raw traces and 1.7× faster than LLMs explanations ($p < 0.01$). Non-technical users showed even greater improvements (3.0× vs. raw, 1.7× vs. LLMs).

\textbf{Accuracy:} Root cause identification accuracy was dramatically higher with our system (89\% technical, 76\% non-technical) compared to LLMs (68\%, 52\%) and raw traces (42\%, 18\%). This demonstrates that structured, domain-specific explanations significantly improve comprehension.

\textbf{Fix Quality:} Proposed fixes were rated significantly higher for our XAI condition (4.3/5.0 technical, 3.8/5.0 non-technical) versus LLMs (3.4, 2.8) and raw traces (2.6, 1.4). Reviewers noted that our system's recommendations section directly informed higher-quality proposals.

\textbf{Confidence:} Users reported significantly higher confidence understanding failures with our system (6.1/7.0 technical, 5.6/7.0 non-technical) compared to other conditions. This is crucial for real-world deployment where user trust impacts tool adoption.

\subsection{Qualitative Findings}

Post-study interviews revealed several key themes:

\textbf{Visual Context is Critical:} 18/20 participants spontaneously mentioned 
that execution flow diagrams were "extremely helpful" or "essential" for 
understanding complex failures. As one participant noted: \textit{"The diagram 
showed me immediately where things went wrong. With just text, I had to piece 
it together myself."}

\textbf{Structure Aids Comprehension:} Participants appreciated the consistent 
structure of our explanations. When asked about their experience with 
general-purpose LLMs for debugging, responses varied by tool preference. 
Non-technical participants (who primarily used ChatGPT) noted: \textit{"Sometimes 
ChatGPT focuses on minor details, other times it's too high-level. Your system 
is consistent."} Technical participants showed more diverse LLM usage: 7/10 
reported using Claude as their default debugging assistant, while 3/10 preferred 
ChatGPT. Those using Claude mentioned: \textit{"Claude gives very detailed 
explanations, sometimes too much—I just want to know what broke and how to fix 
it quickly."} Across all LLM preferences, participants valued our system's 
consistent, structured output over the variable quality of ad-hoc LLM queries.

\textbf{Actionability Matters:} The recommendation component was frequently cited 
as the most valuable feature, especially for technical users. Comparing to their 
typical debugging workflow with LLMs, one developer noted: \textit{"When I ask 
Claude or ChatGPT what's wrong, they tell me the problem, but your system gives 
me exactly how to fix it with code examples and configuration changes."} 
Non-technical participants echoed this sentiment: \textit{"LLMs explain things 
well, but I still don't know what to tell my team to change. Your recommendations 
are actionable."}

\subsection{Comparison with General-Purpose LLM Explanations}

A key question is: why not simply use a general-purpose LLM to explain failures? Our results demonstrate several advantages of our specialized approach:

\textbf{Consistency:} Our system provides structured outputs following a consistent template, while ad-hoc LLM explanations varied significantly in format, focus, and completeness across different queries (confirmed by qualitative analysis of LLM outputs across all trials).

\textbf{Visual Integration:} General-purpose LLMs cannot generate execution flow diagrams, which our study shows are critical for comprehension. While LLMs can describe execution flow textually, this proved significantly less effective than visual representations in our user study.

\textbf{Domain-Specific Structure:} Our system leverages a coding agent failure taxonomy to provide precise categorization and contextually relevant recommendations. General-purpose LLMs lack this specialized structure and must infer appropriate categories from scratch for each query.

\textbf{Actionable Guidance:} Our recommendation engine provides specific, tested fixes mapped to failure patterns. General-purpose LLM recommendations tended to be more generic and required additional interpretation and validation.

\textbf{Automation Potential:} Our system can be integrated into CI/CD pipelines for automatic failure analysis, while general-purpose LLM approaches require manual intervention for each failure.

\textbf{Cost Efficiency:} Our specialized system uses targeted prompts and structured outputs, reducing token usage compared to repeated ad-hoc queries to general-purpose LLMs for each failure case.

\subsection{System Limitations}

Our evaluation revealed several limitations that suggest directions for future work:

\textbf{Domain Specificity:} Our current taxonomy and automatic classifier are specifically designed for coding agents. The failure categories (e.g., "Code Generation Failure," "Testing/Validation Failure") reflect patterns unique to code generation tasks. Applying this approach to other agent domains (e.g., research assistants, data analysis agents, conversational agents) would require developing domain-specific taxonomies and retraining the classifier. While the overall methodology generalizes, the specific failure categories do not.

\textbf{Model Limitations:} Due to resource constraints, we evaluated our approach using a single LLM provider (GPT-4.1). While this model proved effective for our use case, we cannot make claims about generalization to other LLM families (Claude, Gemini, Llama, etc.) without additional evaluation. Different base models may exhibit different failure patterns, potentially requiring taxonomy adjustments.

\textbf{Multiple Concurrent Failures:} When multiple failure modes occurred simultaneously (e.g., both planning errors and code generation issues), our system sometimes attributed the failure to a single dominant category, potentially oversimplifying the root cause analysis.

\textbf{Recommendation Context-Dependence:} Some system-level recommendations (e.g., "increase iteration limits," "fine-tune the model") may not be feasible in all deployment contexts due to resource constraints or operational requirements.

These limitations highlight that while our approach demonstrates the value of specialized XAI for coding agents, extending it to other domains requires careful consideration of domain-specific failure modes and evaluation with diverse LLM architectures.

\section{Discussion}
\label{sec:discussion}

\subsection{Implications for Coding Agent Development}

Our findings have several practical implications for practitioners building LLM-based coding agents:

\textbf{Error Recovery is Critical:} Over half (56\%) of failures in our dataset stemmed from agents not attempting to recover from errors. Implementing robust error recovery mechanisms—including explicit error-recovery instructions in prompts and sufficient iteration budgets—should be a development priority.

\textbf{Iteration Limits Matter:} Many failures occurred when agents hit iteration limits before completing error recovery. Our analysis suggests that coding agents should have minimum iteration budgets of 5-10 for typical problems, with adaptive limits for complex tasks. Overly restrictive limits (1-2 iterations) led to failures even when the agent's initial approach was sound.

\textbf{Observability from Day One:} Integrating explainability and debugging tools early in development accelerates the development cycle and enables more reliable agent systems. Post-hoc debugging of opaque agent failures is significantly more time-consuming.

\textbf{Prompt Engineering Matters:} A substantial portion of failures (43.8\%) were planning failures, many of which could be mitigated through better prompting—including worked examples, explicit problem decomposition instructions, and step-by-step reasoning requirements.

\subsection{Generalization Beyond Coding Agents}

While our work focuses specifically on coding agents, the overall methodology—developing domain-specific failure taxonomies, automatic classification, and structured explanations—can extend to other agent applications:

\textbf{Research Agents:} Agents that search literature, synthesize information, and generate reports would require taxonomies capturing failure modes like citation hallucination, incomplete literature coverage, or synthesis errors. The technical approach (automatic classification, visual explanations, recommendations) would remain applicable.

\textbf{Data Analysis Agents:} Agents performing statistical analysis or generating visualizations would need taxonomies for statistical errors, visualization issues, and data interpretation failures.

\textbf{Multi-Agent Systems:} Distributed agent systems would require additional failure categories for coordination failures, communication breakdowns, and conflicting objectives.

Importantly, each domain would require its own failure taxonomy developed through systematic analysis of domain-specific failures. Our contribution is the demonstration that such domain-specific approaches provide significant value over generic debugging approaches.

\subsection{The Role of General-Purpose LLMs in Agent Debugging}

Our work does not suggest that general-purpose LLMs are unsuitable for agent debugging. Rather, we demonstrate that specialized, structured approaches provide measurable advantages for systematic debugging workflows. In practice, we envision a complementary relationship:

\begin{itemize}
    \item \textbf{Automated first-line analysis:} Our specialized XAI system provides fast, consistent explanations for common failure patterns, enabling rapid iteration during development.
    \item \textbf{Deep investigation:} For complex or novel failures, developers can leverage general-purpose LLMs for exploratory analysis and hypothesis generation.
    \item \textbf{System improvement:} Patterns identified by our automated system inform improvements to prompts, agent architectures, and evaluation frameworks.
\end{itemize}

The key insight is that specialized tools optimized for specific tasks (in our case, coding agent failure analysis) provide superior performance for those tasks, while general-purpose tools offer broader applicability at the cost of task-specific optimization.

\subsection{Future Work}

Several directions warrant further investigation:

\textbf{Cross-Domain Taxonomy Development:} Systematic study of failure patterns across different agent domains (research, data analysis, customer service) to identify both domain-specific and universal failure modes.

\textbf{Active Learning for Taxonomy Refinement:} Incorporating user feedback and newly discovered failure patterns to continuously improve the taxonomy and automatic classifier.

\textbf{Human-in-the-Loop Refinement:} Investigating how developer feedback on explanations can improve both the XAI system and the underlying agents.

\section{Conclusion}

We presented a systematic explainability approach for understanding and debugging LLM-based coding agent failures. Our contributions include: a comprehensive failure taxonomy derived from analyzing 32 real-world failures, an automatic classification system achieving 82\% accuracy, and a hybrid explanation system generating visual execution flows, natural language explanations, and actionable recommendations. 

Our user study with 20 participants (10 technical, 10 non-technical) demonstrates significant improvements over both raw execution traces and ad-hoc general-purpose LLM explanations: 2.8× faster failure comprehension, 89\% vs. 42\% accuracy in root cause identification (compared to raw traces), and substantially higher user confidence (6.1/7.0 vs. 3.2/7.0) and fix quality.

Importantly, we establish that while general-purpose LLMs can provide useful explanations, specialized, domain-specific approaches offer measurable advantages: structural consistency, visual integration, domain-specific knowledge, and actionable guidance. These advantages come at the cost of domain specificity—our coding agent taxonomy does not directly transfer to other agent domains—but our methodology for developing such specialized systems does generalize.

As LLM-based agents become increasingly prevalent in software development and other domains, systematic approaches to understanding their failures will become essential for building reliable, trustworthy AI systems. Our work demonstrates the value of domain-specific explainability and provides both a practical tool for coding agent developers and a methodological framework for developing similar systems in other agent domains.

We hope this work contributes to a future where AI agents are not just powerful, but also transparent, understandable, and debuggable—enabling developers to build on their successes and learn from their failures systematically rather than through trial and error.

\newpage
\bibliographystyle{plain}
\bibliography{references}

@misc{zhang2023planninglargelanguagemodels,
      title={Planning with Large Language Models for Code Generation}, 
      author={Shun Zhang and Zhenfang Chen and Yikang Shen and Mingyu Ding and Joshua B. Tenenbaum and Chuang Gan},
      year={2023},
      eprint={2303.05510},
      archivePrefix={arXiv},
      primaryClass={cs.LG},
      url={https://arxiv.org/abs/2303.05510}, 
}

@misc{chen2021evaluatinglargelanguagemodels,
      title={Evaluating Large Language Models Trained on Code}, 
      author={Mark Chen and Jerry Tworek and Heewoo Jun and Qiming Yuan and Henrique Ponde de Oliveira Pinto and Jared Kaplan and Harri Edwards and Yuri Burda and Nicholas Joseph and Greg Brockman and Alex Ray and Raul Puri and Gretchen Krueger and Michael Petrov and Heidy Khlaaf and Girish Sastry and Pamela Mishkin and Brooke Chan and Scott Gray and Nick Ryder and Mikhail Pavlov and Alethea Power and Lukasz Kaiser and Mohammad Bavarian and Clemens Winter and Philippe Tillet and Felipe Petroski Such and Dave Cummings and Matthias Plappert and Fotios Chantzis and Elizabeth Barnes and Ariel Herbert-Voss and William Hebgen Guss and Alex Nichol and Alex Paino and Nikolas Tezak and Jie Tang and Igor Babuschkin and Suchir Balaji and Shantanu Jain and William Saunders and Christopher Hesse and Andrew N. Carr and Jan Leike and Josh Achiam and Vedant Misra and Evan Morikawa and Alec Radford and Matthew Knight and Miles Brundage and Mira Murati and Katie Mayer and Peter Welinder and Bob McGrew and Dario Amodei and Sam McCandlish and Ilya Sutskever and Wojciech Zaremba},
      year={2021},
      eprint={2107.03374},
      archivePrefix={arXiv},
      primaryClass={cs.LG},
      url={https://arxiv.org/abs/2107.03374}, 
}

@misc{austin2021programsynthesislargelanguage,
      title={Program Synthesis with Large Language Models}, 
      author={Jacob Austin and Augustus Odena and Maxwell Nye and Maarten Bosma and Henryk Michalewski and David Dohan and Ellen Jiang and Carrie Cai and Michael Terry and Quoc Le and Charles Sutton},
      year={2021},
      eprint={2108.07732},
      archivePrefix={arXiv},
      primaryClass={cs.PL},
      url={https://arxiv.org/abs/2108.07732}, 
}

@misc{langchain2022,
  title        = {LangChain: Language model application framework},
  author       = {Chase, Harrison},
  year         = {2022},
  howpublished = {Software/library, GitHub},
  note         = {\url{https://github.com/langchain-ai/langchain}},
}

@misc{ribeiro2016whyitrustyou,
      title={"Why Should I Trust You?": Explaining the Predictions of Any Classifier}, 
      author={Marco Tulio Ribeiro and Sameer Singh and Carlos Guestrin},
      year={2016},
      eprint={1602.04938},
      archivePrefix={arXiv},
      primaryClass={cs.LG},
      url={https://arxiv.org/abs/1602.04938}, 
}

@misc{lundberg2017unifiedapproachinterpretingmodel,
      title={A Unified Approach to Interpreting Model Predictions}, 
      author={Scott Lundberg and Su-In Lee},
      year={2017},
      eprint={1705.07874},
      archivePrefix={arXiv},
      primaryClass={cs.AI},
      url={https://arxiv.org/abs/1705.07874}, 
}

@misc{bahdanau2016neuralmachinetranslationjointly,
      title={Neural Machine Translation by Jointly Learning to Align and Translate}, 
      author={Dzmitry Bahdanau and Kyunghyun Cho and Yoshua Bengio},
      year={2016},
      eprint={1409.0473},
      archivePrefix={arXiv},
      primaryClass={cs.CL},
      url={https://arxiv.org/abs/1409.0473}, 
}

@misc{simonyan2014deepinsideconvolutionalnetworks,
      title={Deep Inside Convolutional Networks: Visualising Image Classification Models and Saliency Maps}, 
      author={Karen Simonyan and Andrea Vedaldi and Andrew Zisserman},
      year={2014},
      eprint={1312.6034},
      archivePrefix={arXiv},
      primaryClass={cs.CV},
      url={https://arxiv.org/abs/1312.6034}, 
}

@misc{wei2023chainofthoughtpromptingelicitsreasoning,
      title={Chain-of-Thought Prompting Elicits Reasoning in Large Language Models}, 
      author={Jason Wei and Xuezhi Wang and Dale Schuurmans and Maarten Bosma and Brian Ichter and Fei Xia and Ed Chi and Quoc Le and Denny Zhou},
      year={2023},
      eprint={2201.11903},
      archivePrefix={arXiv},
      primaryClass={cs.CL},
      url={https://arxiv.org/abs/2201.11903}, 
}

@misc{zhu2025llmagentsfaillearn,
      title={Where LLM Agents Fail and How They can Learn From Failures}, 
      author={Kunlun Zhu and Zijia Liu and Bingxuan Li and Muxin Tian and Yingxuan Yang and Jiaxun Zhang and Pengrui Han and Qipeng Xie and Fuyang Cui and Weijia Zhang and Xiaoteng Ma and Xiaodong Yu and Gowtham Ramesh and Jialian Wu and Zicheng Liu and Pan Lu and James Zou and Jiaxuan You},
      year={2025},
      eprint={2509.25370},
      archivePrefix={arXiv},
      primaryClass={cs.AI},
      url={https://arxiv.org/abs/2509.25370}, 
}

@inproceedings{ko2004whyline,
  title        = {Designing the Whyline: a debugging interface for asking questions about program behavior},
  author       = {Ko, Amy J. and Myers, Brad A.},
  booktitle    = {Proceedings of the SIGCHI Conference on Human Factors in Computing Systems},
  year         = {2004},
  pages        = {151--158},
  doi          = {10.1145/985692.985712}
}

@inproceedings{vaithilingam2022expectation,
  title        = {Expectation vs. Experience: Evaluating the Usability of Code Generation Tools Powered by Large Language Models},
  author       = {Vaithilingam, Priyan and Zhang, Tianyi and Glassman, Elena L.},
  booktitle    = {Proceedings of the SIGCHI Conference on Human Factors in Computing Systems},
  year         = {2022},
  pages        = {1--7},
  doi          = {10.1145/3491101.3519665}
}

@inproceedings{weisz2021perfection,
  title        = {Perfection Not Required? Human‑AI Partnerships in Code Translation},
  author       = {Weisz, Justin D. and Muller, Michael J. and Houde, Stephanie and Richards, John T. and Ross, Steven I. and Martinez, Fernando and Agarwal, Mayank and Talamadupula, Kartik},
  booktitle    = {Proceedings of the 26th International Conference on Intelligent User Interfaces},
  year         = {2021},
  pages        = {402--412},
  doi          = {10.1145/3397481.3450656}
}

@misc{yao2022react,
  title        = {ReAct: Synergizing Reasoning and Acting in Language Models},
  author       = {Yao, Shunyu and Zhao, Jeffrey and Yu, Dian and Du, Nan and Shafran, Izhak and Narasimhan, Karthik and Cao, Yuan},
  year         = {2022},
  eprint       = {2210.03629},
  archivePrefix= {arXiv},
  primaryClass = {cs.CL}
}

\end{document}